%% file: QM14-template.tex
\documentclass[3p,times]{elsarticle}

\usepackage{ecrc}

\usepackage{epstopdf}

\volume{00}

\firstpage{1}

\journalname{Nuclear Physics A}

\runauth{Leonardo Milano for the ALICE Collaboration}

\jid{nupha}

\jnltitlelogo{Nuclear Physics A}

\usepackage{graphicx}
\usepackage{amsmath,amssymb}
\usepackage{hyperref}
\usepackage{amssymb}
\usepackage{amsthm}

\graphicspath{{./img/}}

\usepackage{graphicx}
\usepackage{dcolumn}   
\usepackage{bm}        
\usepackage{amssymb}   
\usepackage{amsfonts}
\usepackage{graphics}
\usepackage{grffile}   
\usepackage{epsfig}
\usepackage{units}
\usepackage[usenames]{color}
\usepackage[normalem]{ulem} 
\usepackage{color}
\usepackage[utf8]{inputenc}
\usepackage[T1]{fontenc}
\usepackage{lineno}
\usepackage{wrapfig}
\modulolinenumbers[5]
\usepackage{lineno}
\input{commands.tex}

\begin{document}

\begin{frontmatter}



\title{Long-range angular correlations at the LHC with ALICE}

\author{Leonardo Milano (for the ALICE\fnref{col1} Collaboration)}
\fntext[col1] {A list of members of the ALICE Collaboration and acknowledgements can be found at the end of this issue.}
\address{CERN, the European Organisation for Nuclear Research

\href{mailto:Leonardo.Milano@cern.ch}{Leonardo.Milano@cern.ch}}




\begin{abstract}
The observation of long-range correlations on the near and away side ---the double ridge--- in high-multiplicity \pPb\ collisions at \snn = 5.02 TeV and its similarity to \PbPb\ collisions remains one of the open questions from the \pPb\ run at the Large Hadron Collider (LHC).
It has been attributed to mechanisms that involve initial-state effects, such as gluon saturation and colour connections forming along the longitudinal direction, and final-state effects, such as parton-induced interactions and collective effects developing in a high-density system possibly formed in these collisions.
In order to understand the nature of this double-ridge structure the two-particle correlation analysis has been extended to identified particles.
The observed mass dependence in p-Pb resembles the pattern observed in \PbPb\ collisions and is compatible with expectations from hydrodynamics. The subtraction of the long-range structures allows the study of the hadron production belonging to the fragmentation of jets originating from low momentum-transfer scatterings (minijets).

\end{abstract}

\begin{keyword}
two-particle correlations
\sep elliptic flow
\sep particle identification 
\sep ridge
\sep cumulants
\sep collective flow
\sep ALICE
\sep LHC
\sep multiparton interactions


\end{keyword}

\end{frontmatter}



\section{Introduction}
\label{intro} 
The study of particle correlations is a powerful tool to probe the mechanism of particle production in collisions of hadrons and nuclei at high beam energy.
The correlation of particles is measured in relative angles $\Dphi$ and $\Deta$, where $\Dphi$ and $\Deta$ are the differences in azimuthal angle~$\varphi$ and pseudorapidity~$\eta$ between two particles.
The observation of the ridge structures in high-multiplicity events on the near ($\Dphi \approx 0$) and away side ($\Dphi \approx \pi$) of the trigger particle, extending over a long-range in \Deta\ is one of the unexpected observations of the \pPb\ run at the LHC \cite{CMSridge,ATLASridge,ALICEridge}.
The excellent particle identification capability of the ALICE detector \cite{ALICEdetector} allows to further characterise the system created in high-multiplicity \pPb\ collisions in order to understand the origin of this effect. Multi-particle correlations are also studied to further reduce the contribution mainly originating from the fragmentation of hard partons. After subtraction of the long-range structures the hadron production belonging to the fragmentation of jets originating from low momentum-transfer scatterings (minijets) can be studied.
\section{Results}
\label{results} 
The correlation between a trigger hadron and identified pions, kaons and
protons\footnote{Pions, kaons and protons, as well as the symbols $\pi$, K and p, refer to the sum of particles and antiparticles.} has been studied in order to explore the mass dependence of the second-order coefficient $v_2$ extracted from two-particle correlations \cite{ALICEPIDridge}.
The correlation is expressed in terms of the associated yield per trigger particle where both particles are from the same transverse momentum \pt interval in a fiducial region of $|\eta|<0.8$:
\begin{equation}
  \frac{1}{\Ntrig} \dNassoc = \frac{S(\Deta,\Dphi)}{B(\Deta,\Dphi)} \label{pertriggeryield}
\end{equation}
where \Ntrig\ is the total number of trigger particles in the event class and \pt interval.
The signal distribution $S(\Deta,\Dphi) = 1/\Ntrig\ \dd^2N_{\rm same}/\dd\Deta\dd\Dphi$ is the associated yield per trigger particle for particle pairs from the same event.
The background distribution $B(\Deta,\Dphi) = \alpha\ \dd^2N_{\rm  mixed}/\dd\Deta\dd\Dphi$ is constructed by correlating the trigger particles in one event with the associated particles from other events of the same event class and within the same \unit[2]{cm}-wide $z_{\rm vtx}$ interval and corrects for pair acceptance and pair efficiency.
Events are divided in fractions of the analysed event sample, ordered according
to the charge deposition in the \VZEROA\ detector, and denoted “0–20\%”, “20–40\%”, “40–60\%”, “60–100\%” from the highest to the lowest multiplicity.
Particles are identified by means of the specific energy loss in the gas of the Time Projection Chamber (\TPC) and the arrival time in the Time Of Flight (\TOF) detector.
Assuming the low-multiplicity class to be dominated by correlations from fragmentation of hard partons, it is possible to reduce the jet contribution by subtracting the correlation function of the low-multiplicity class from the high-multiplicity class. This procedure was applied for the first time in Ref.\cite{ALICEridge} and widely used to study the \pPb\ system, where no significant evidence of energy loss of partons in the medium is observed \cite{ALICERpPb}.

\begin{figure}[ht!f]
  \centering
  \includegraphics[width=.9\textwidth]{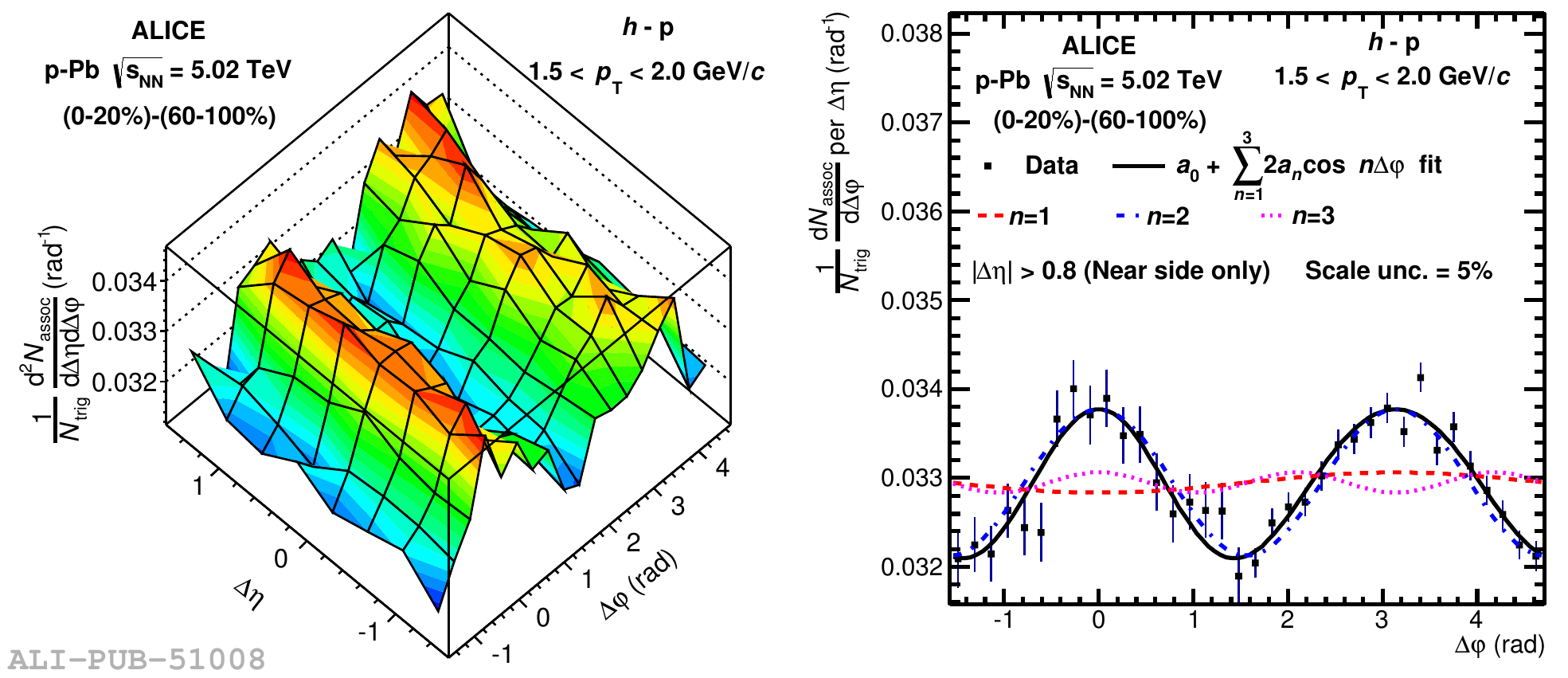}
  \caption{\label{fig:pertriggeryields_subtracted}
    Left panel: associated yield per trigger particle as a function of $\Dphi$ and $\Deta$ for $h-{\rm p}$ correlations for $1.5<\pt<$~\unit[2]{\gevc} for the 0--20\% event class where the corresponding correlation from the 60--100\% event class has been subtracted. Right panel: projection of the left panel to $\Dphi$ averaged over $0.8 < |\Deta| < 1.6$ on the near side and $|\Deta| < 1.6$ on the away side. The figure contains only statistical uncertainty. Systematic uncertainties are mostly correlated and are less than 5\%.
  }
\end{figure}

In the left panel of Fig.~\ref{fig:pertriggeryields_subtracted} the resulting $h-{\rm p}$ correlation for $1.5<\pt<$~\unit[2]{\gevc} is shown.
The projection onto $\Dphi$ is averaged over $0.8 < |\Deta| < 1.6$ on the near side and $|\Deta| < 1.6$ on the away side.
From the relative modulations (for a particle species $i$) $V_{n\Delta}^{h-i}\{{\rm 2PC, sub}\} = a_n^{h-i} / (a_0^{h-i}+b)$, where $a_{n}^{h-i}$ is the $a_{n}$ extracted from $h-i$ correlations and $b$ is the combinatorial baseline of the lower-multiplicity class which has been subtracted,   the $v_n\{{\rm 2PC,sub}\}$ coefficient of order $n$ can be extracted as explained in \cite{ALICEPIDridge}.

\begin{figure}[ht!f]
  \centering
  \includegraphics[width=0.84\textwidth]{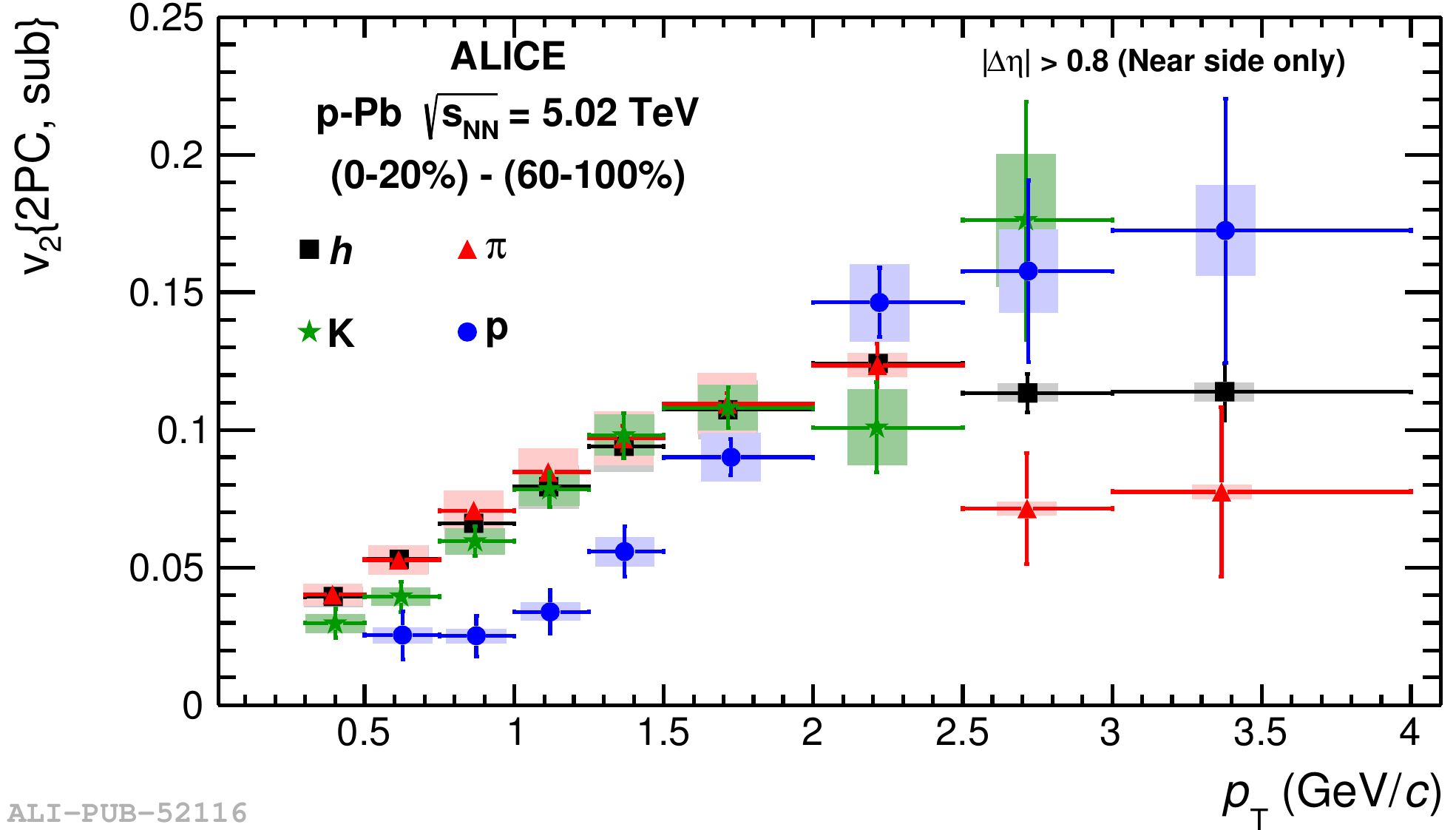}
  \caption{\label{fig:v2_subtracted}
    The Fourier coefficient $v_2\{{\rm 2PC,sub}\}$ for hadrons (black squares), pions (red triangles), kaons (green stars) and protons (blue circles) as a function of $\pt$ from the correlation in the 0--20\% multiplicity class after subtraction of the correlation from the 60--100\% multiplicity class. The data are plotted at the average-$\pt$ for each considered $\pt$ interval and particle species under study. Error bars show statistical uncertainties while shaded areas denote systematic uncertainties.
  }
\end{figure}

Figure~\ref{fig:v2_subtracted} shows the extracted $v_2\{{\rm 2PC,sub}\}$ coefficients for $h$, $\pi$, K and p as a function of $\pt$.
The coefficient $v_2^{\rm p}$ is significantly lower than $v_2^\pi$ for $0.5 < \pt <$~\unit[1.5]{\gevc}, and larger than $v_2^\pi$ for $\pt >$~\unit[2.5]{\gevc}. The crossing occurs at $\pt \approx$ \unit[2]{\gevc}. The coefficient $v_2^{\rm K}$ is consistent with $v_2^\pi$ above \unit[1]{\gevc}; below \unit[1]{\gevc} there is a hint that $v_2^{\rm K}$ is lower than $v_2^\pi$.

The mass ordering and crossing is a particularly interesting observation because it is qualitatively similar to observations in nucleus--nucleus collisions  \cite{ALICEPIDflow}. This pattern is in agreement with the expectations from hydrodynamic models in \pPb\ \cite{ModelWernerHidropPb,ModelBozekHydropPb}.

The analysis of multi-particle cumulants shows that the $v_2$ coefficient has a finite value when calculated with two- and four-particle cumulants. This observation suggests that the long-range double-ridge structure arises from a correlation among many particles \cite{ALICECumulantsQM}.

The qualitative similarity between proton--nucleus and nucleus--nucleus extends to other observables, such as the multiplicity dependence of the baryon-to-meson ratio and the mean transverse momentum <\pt> \cite{ALICEPIDspectra}.

To have a more complete picture of the physical phenomena involved in \pPb\ collisions, it is interesting to study the fragmentation of partons from hard processes in the \pt range where these ridge-like structures have been observed.
In order to isolate the jet-like hadron production in this low-\pt region, the long-range pseudorapidity ridge-like structures are subtracted.
If the same \pt selection for associated and trigger particle is used, it is possible to experimentally define the number of uncorrelated seeds as:
\begin{equation}
  <N_{\mathrm{uncorrelated~seeds}}>=\frac{<N_{\mathrm{triggers}}>}{<N_{\mathrm{correlated~triggers}}>}=\frac{<N_{\mathrm{triggers}}>}{1+<N_{\mathrm{associated,~near~side}}>+<N_{\mathrm{associated,~away~side}}>}
\end{equation}
In PYTHIA, for \pp\ collisions \cite{ALICEMPIpp}, the uncorrelated seeds are found to be linearly correlated to the number of MPIs in a certain \pt range, independent of the $\eta$ range explored.

\begin{figure}[ht!f]
  \centering
  \includegraphics[width=0.77\textwidth]{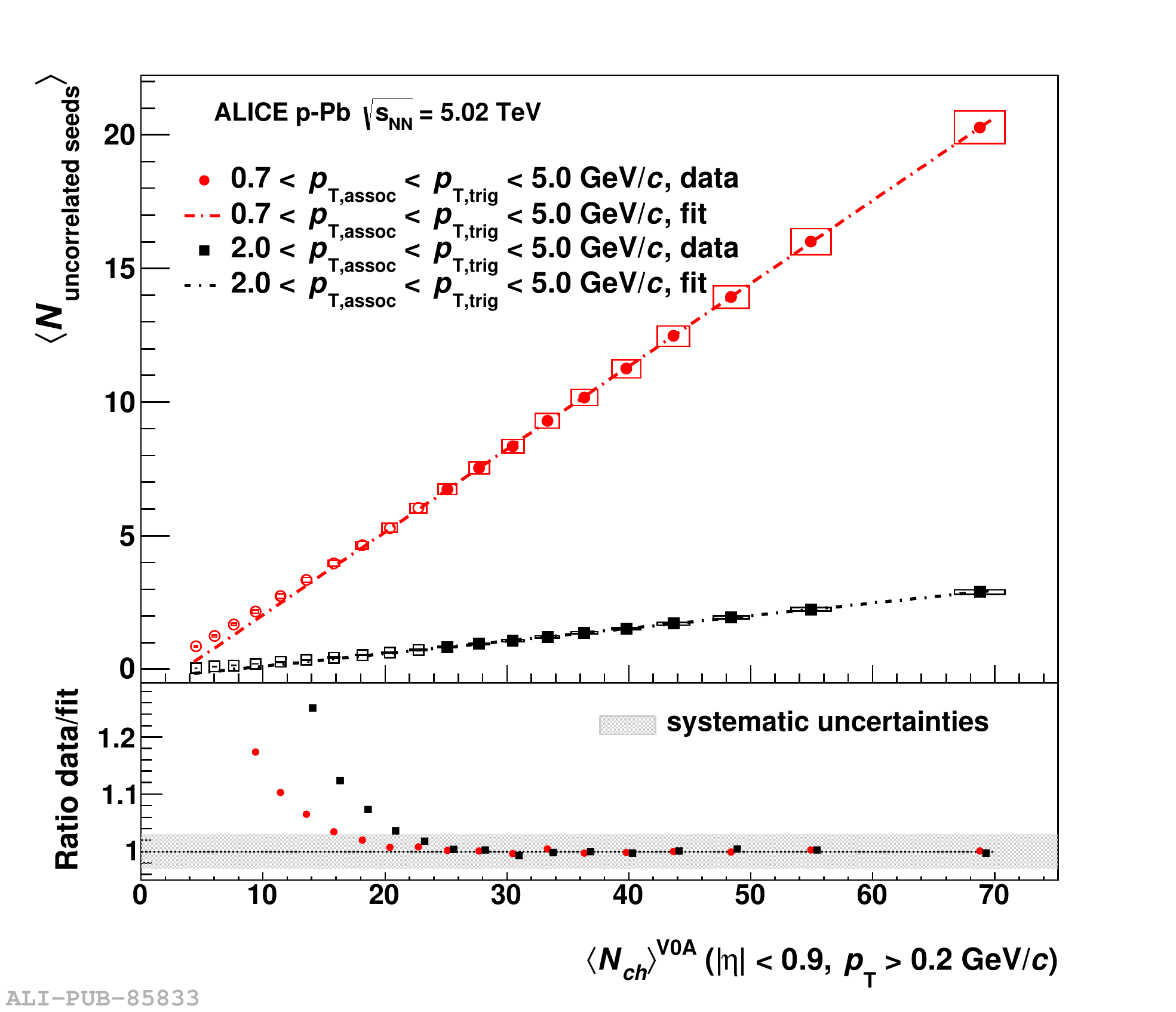}
  \caption{\label{fig:MPI}
Top panel:number of uncorrelated seeds as a function of the midrapidity charged particle multiplicity. Shown are results for two \pt cuts. Each of them is fit with a linear function in the 0–50\% multiplicity classes; open symbols are not included in the fit. Statistical (lines) and systematic uncertainties (boxes) are shown, even though the statistical ones are smaller than the symbol size. Bottom panel: ratio between the number of uncorrelated seeds and the linear fit functions. Black points are displaced slightly for better visibility.}\end{figure}

Figure \ref{fig:MPI} presents the uncorrelated seeds as a function of the midrapidity charged-particle multiplicity $N_{Ch}$ for two \pt cuts.
The uncorrelated seeds exhibit a linear increase with $N_{Ch}$ in particular at high multiplicity in \pPb\ collisions. To quantify this behaviour, a linear fit is performed in the 0–50\% multiplicity class and the ratio to the data is presented in the bottom panel. The linear description extends over a wide range ($N_{Ch} \gtrsim 20$), with a departure from linear behaviour observed at low multiplicity. This observation suggests that there is no evidence of a saturation in the number of multiple parton interactions in high-multiplicity \pPb \cite{ALICEMPIpPb}, compared to \pp\ where it deviates from the linear dependence at large $N_{Ch}$ values \cite{ALICEMPIpp}.
\section{Conclusions}
The experimental observations in high-multiplicity \pPb\ collisions are highly suggestive of collective behaviour. The $v_2$ coefficient extracted from two-particle correlations shows a mass ordering that is reminiscent of the observations in \PbPb\ collisions, where a collective behaviour of the system is established. The analysis of multi-particle correlations suggests that the double ridge in high-multiplicity \pPb\ collisions arises from a correlations among many particles. A further hint of similarity between the two systems is found when looking at the multiplicity dependence of the particle ratios and <\pt>.
The study of the minijet production in \pPb\ indicates that events in high-multiplicity \pPb\ collisions can be modelled as an incoherent fragmentation of multiple parton–parton scatterings. 
Future studies of two-particle correlations with very large separation in pseudorapidity in \pPb\ will give the opportunity to further probe the low-\textit{x} regime where saturation effects are expected to be stronger and to study the presence of the same effect in smaller systems, such as \pp.







\end{document}

%% file: commands.tex
\newcommand{\gevc}         {GeV/\ensuremath{c}}

\newcommand{\TOF}          {\rm{TOF}}

\newcommand{\TPC}          {\rm{TPC}}

\newcommand{\VZEROA}       {\rm{V0A}}

\newcommand{\pp}           {pp}

\newcommand{\PbPb}         {\mbox{Pb--Pb}}

\newcommand{\pPb}          {\mbox{p--Pb}}

\newcommand{\pt}           {\ensuremath{p_{\mathrm{T}}}{ }}

\newcommand{\snn}          {\ensuremath{\sqrt{s_{\mathrm{NN}}}}}

\newcommand{\dd}           {\ensuremath{\mathrm{d}}}

\newcommand{\Dphi}         {\ensuremath{\Delta\varphi}}
\newcommand{\Deta}         {\ensuremath{\Delta\eta}}
\newcommand{\Ntrig}        {\ensuremath{N_{\mathrm{trig}}}}

\newcommand{\dNassoc}      {\ensuremath{\frac{\dd^2N_{\mathrm{assoc}}}{\dd\Deta\dd\Dphi}}}

\newcommand{\com}[1]       {}

%% file: QM14-template.bbl
\begin{thebibliography}{00}


\bibitem{CMSridge} CMS, Phys. Lett. B 718 (2013) 795
\bibitem{ATLASridge} ATLAS, Phys.Rev.Lett. 110 (2013) 182302
\bibitem{ALICEridge} ALICE, Phys.Lett. B719 (2013) 29–41
\bibitem{ALICEdetector} ALICE, J. Phys. G: Nucl. Part. Phys. (2004) 30 1517
\bibitem{ALICEPIDridge} ALICE, Physics Letters B 726 (2013) 164-177
\bibitem{ALICERpPb} ALICE, arXiv:1405.2737 [nucl-ex]
\bibitem{ALICEPIDflow} ALICE, arXiv:1405.4632 [nucl-ex]
\bibitem{ModelWernerHidropPb} K.Werner, M. Bleicher, B. Guiot, I. Karpenko,T. Pierog, Phys. Rev. Lett. 112, 232301 (2014)
\bibitem{ModelBozekHydropPb} P. Bozek, W. Broniowski, G.Torrieri, Phys. Rev. Lett. 111, 172303 (2013)
\bibitem{ALICECumulantsQM} ALICE,  arXiv:1406.2474 [nucl-ex]
\bibitem{ALICEPIDspectra} ALICE, Physics Letters B 728 (2014) 25–38
\bibitem{ALICEMPIpp}  ALICE, JHEP 09 (2013) 049
\bibitem{ALICEMPIpPb}  ALICE, arXiv:1406.5463 [nucl-ex]

\end{thebibliography}
